\documentclass[conference,10pt]{IEEEtran}
 
\usepackage{cite}
\usepackage{amsmath,amssymb,amsfonts}
\usepackage{algorithmic}
\usepackage{graphicx}
\usepackage{textcomp}
\usepackage{amsmath}
\usepackage{bm}
\usepackage{amsmath,bbm,amssymb,amsfonts,amstext,amsopn}
\usepackage{amsthm}
\usepackage{amssymb}
\usepackage{epstopdf}

\usepackage{amsthm}
\usepackage{psfrag}	

\usepackage{amsmath,bbm,amssymb,amsfonts,amstext,amsopn}
\DeclareMathOperator{\Tr}{Tr}
\DeclareMathOperator{\Rank}{Rank}

\allowbreak
\usepackage{url}
\usepackage{epsfig}
\usepackage{setspace}
\usepackage{multirow}
\usepackage{float}
\usepackage{amsthm}

\newcommand{\subparagraph}{}
\usepackage[english]{babel}
\usepackage{amsthm}
\theoremstyle{remark}

\usepackage{etoolbox}
\usepackage{algorithm}
\usepackage{epsfig}
\usepackage{caption}
\usepackage{algorithmic}
\usepackage{xcolor}

\allowdisplaybreaks
  \makeatletter
\newcommand{\linebreakand}{%
\end{@IEEEauthorhalign}
\hfill\mbox{}\par
\mbox{}\hfill\begin{@IEEEauthorhalign}
}
\makeatother
\begin{document}
	\captionsetup[figure]{labelformat={default},labelsep=period,name={Fig.}}
\title{Resource Allocation for IRS-Enabled Secure Multiuser Multi-Carrier 
Downlink URLLC Systems}
\author{\IEEEauthorblockN{Mohammad Naseri Tehrani} 
\IEEEauthorblockN{Friedrich-Alexander-University (FAU),\\ 
Erlangen-Nuremberg, Germany.\\ 
email:~\emph{moh.naseritehrani@fau.de}}\and 
\IEEEauthorblockN{Shahrokh Farahmand} 
\IEEEauthorblockA{Iran 
University of 
Science and Technology (IUST),\\ Tehran, Iran. \\ email:\emph{ 
shahrokhf@iust.ac.ir}} }
\maketitle \vspace*{-5mm}
\begin{abstract}
Secure ultra-reliable low-latency communication (URLLC) has been recently
investigated with the fundamental limits of finite block length (FBL) regime in 
mind. Analysis has revealed that when eavesdroppers outnumber BS 
antennas or enjoy a more favorable channel condition compared to the legitimate 
users, base station (BS) transmit power should increase exorbitantly to meet 
quality of 
service (QoS) constraints. Channel-induced impairments such as shadowing and/or 
blockage pose a similar challenge. These practical considerations can 
drastically limit secure URLLC performance in FBL regime. Deployment of 
an intelligent 
reflecting surface (IRS) can endow such systems with much-needed 
\emph{resiliency} and \emph{robustness} to satisfy stringent latency, 
availability, and reliability requirements. We 
address this problem and propose a 
joint design of IRS platform and secure URLLC network. We minimize 
the total BS transmit power by simultaneously designing the beamformers and 
artificial noise at the BS and phase-shifts at the IRS, while guaranteeing the 
required number of securely transmitted bits with the desired packet error 
probability, information leakage, and maximum affordable delay. The proposed 
optimization problem is non-convex and we apply block coordinate descent and 
successive convex approximation to iteratively solve a series of convex 
sub-problems instead. The proposed 
algorithm converges to a sub-optimal solution in a few iterations and attains 
substantial power saving and robustness compared to baseline schemes.   
\end{abstract}  

\section{Introduction} 
Ultra-reliable low-latency communication (URLLC) is founded on two conflicting 
features of high reliability, e.g., bit error 
rates (BERs) of $10^{-6}$, and low latency, e.g., delays of at most 
$1$ms\cite{popski}. In a similar 
fashion, 
physical layer security (PLS) stands out as a promising approach to enhance 
both secrecy and service availability by exploiting the physical 
characteristics of wireless channel. PLS-based resource allocation has relied 
on secrecy capacity formula that 
is valid in the infinite block length regime and under additive white Gaussian 
noise (AWGN) channel assumption \cite{ng2012energy,misoyan}. These resource 
allocation schemes were developed without considering the 
crucial low latency requirement of URLLC users, which is realized by short 
packet transmissions (SPT). To fill this gap, \cite{ghanem2020resourcesec} 
investigated a secure URLLC multiuser downlink setup with a single carrier 
with multiple eavesdroppers. Still, unfavorable 
channel conditions, such as multipath fading, blockage, and spatial correlation 
between BS-users and BS-eavesdroppers channels, would severely affect QoS, 
energy efficiency, and security\cite{ghanem2020resourcesec}. Subsequently, the 
required secure number of bits can not be guaranteed at the intended receiver. 

In this regard, IRS-assisted 
communication has enhanced the performance of 
different communication techniques such as multi-carrier transmissions 
\cite{yang2020intelligent}, multi-antenna 
communications \cite{zhang2020capacity}, and PLS \cite{cui2019secure}. Most of 
previous works in IRS resource allocation mainly focused on the single-carrier 
communications \cite{kammoun2020asymptotic}. However, multi-carrier 
communications provides a host of desirable features such as simplified 
equalization, multi-user diversity and flexible resource allocation of power 
and bandwidth. As a challenge for multi-carrier techniques, IRS reflection 
coefficients need to be designed to serve all sub-carriers efficiently and 
simultaneously \cite{yang2020intelligent}. Only 
recently, URLLC resource allocation has begun to benefit from the advantages 
IRSs offer \cite{ghanem2020joint}. However, 
a joint investigation of resource allocation for secure multi-carrier URLLC 
when IRS is deployed is missing from the literature. 

Targeting this research gap, our work's main contribution is to study the 
problem of minimizing the total BS transmit power by jointly designing the 
beamformers and artificial noise (AN) at the BS and phase shifts at the IRS, 
subject to a required minimum secure rate of URLLC users in the finite block 
length regime. Compared to existing literature on secure URLLC, this work is 
different from \cite{ghanem2020resourcesec} as it 
is both multi-carrier and employs an IRS. It is different from 
\cite{ren2020resource} as \cite{ren2020resource} is both single carrier and 
single antenna and solves the loosely speaking dual problem of maximizing 
sum-secure-rate subject to latency and power constraints. Furthermore, there is 
no IRS in \cite{ren2020resource}. The posed problem is non-convex with strong 
coupling between design variables. To address these challenges, we leverage 
optimization techniques such as block coordinate descent (BCD) and successive 
convex approximation (SCA). Instead of relaxing the ensuing sub-problems by 
dropping rank constraints which may render the obtained solution infeasible, we 
utilize an iterative penalty-based SCA method that transforms the rank 
constraint into linear matrix inequalities (LMIs) per iteration \cite{dai19}.


\section{System Model}
\begin{figure}
 	\centering
 	\scalebox{0.80}{
 		\includegraphics[width=1.2\linewidth]{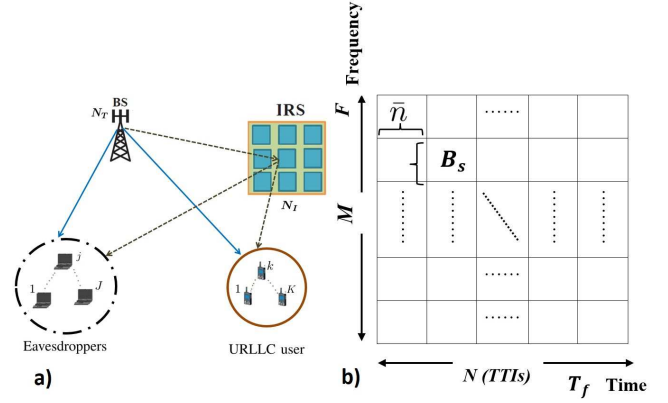}	
 	}
 	\caption{System setup and model parameters}
 	\label{model}
 	\vspace{-0.5cm}
 \end{figure}

We consider a single cell in downlink mode, where a BS equipped with $N_{T}$ 
antennas is trying to transmit data to $K$ single-antenna URLLC users indexed 
by $k =\{1,\dots,K\}$. There exist $J$ single-antenna eavesdroppers indexed by 
$j =\{1,\dots,J\}$. An IRS with $N_{I}$ elements is deployed to help the BS 
communicate securely with the intended URLLC users, cf. Fig.~\ref{model}. We 
use frames of duration of $T_{f}$ seconds, where each frame 
is divided into $N$ time slots indexed by $n =\{1,\dots,N\}$. $\bar{n}$ symbols 
are transmitted during each time slot. Total bandwidth $F$ is divided into 
$M$ sub-carriers with $B_s:=F/M$ Hz of bandwidth each. The value of 
$\bar{n}$ depends on the sub-carrier bandwidth $B_s$ and the total frame 
duration $T_{f}$, 
i.e., $\bar{n}=\frac{B_s T_{f}}{N}$, which is assumed to be an integer value. 
We further assume that the maximum tolerable delay for each user is known at 
the BS and only users whose delay constraint can be satisfied in the 
current frame are admitted.

Upon applying linear beamforming at the BS, signal vector transmitted by BS on 
sub-carrier $m$ in time slot $n$  becomes
 \begin{align}\label{txvector}
 \mathbf{x}[m,n]&=\sum_{k=1}^{K}\mathbf{w}_{k}[m,n]s_{k}[m,n] +\mathbf{v}[m,n],
 \end{align}
 where $\mathbf{w}_{k}[m,n] \in \mathbb{C}^{N_{T} \times 1}$ denotes the 
 beamforming vector for user $k$ on sub-carrier $m$ in time slot $n$, and 
 $s_{k}[m,n] \in \mathbb{C}$ represents independent and identically distributed 
 complex zero-mean, unit variance symbol transmitted to user $k$ on sub-carrier 
 $m$ in time slot $n$. Moreover, $\mathbf{v}[m,n]$ describes the AN component 
 and is modeled as a zero-mean complex circularly-symmetric Gaussian random 
 vector with Hermitian symmetric covariance 
 matrix $\mathbf{V}[m,n]$. We 
 assume a block fading 
 channel model whose coherence time exceeds $T_{f}$. Furthermore, each 
 sub-carrier's bandwidth is smaller than channel coherence bandwidth leading 
 to a flat-fading model. We 
 further assume that perfect channel state information (CSI) is 
 available at the BS. As a result, our proposed algorithm will provide a 
 performance benchmark on any method derived under partial or no CSI 
 availability. The signal 
 received at the $k$-th user is given by
 \begin{align}\label{channeld}
 \hspace{-3mm}y_{k}[m,n]=\bar{\mathbf{h}}^H_k[m]\mathbf{x}[m,n]  + {z}_{k}[m,n],
 \end{align}
where we have defined 
$\bar{\mathbf{h}}^H_k[m]:=\mathbf{h}^{H}_{k}[m]\boldsymbol{\Phi}[n]
\mathbf{H}[m]+\mathbf{g}^{H}_{k}[m]$ and $\mathbf{h}^H_{k}[m] \in \mathbb{C}^{1 
\times N_I}$, $\mathbf{g}^H_{k}[m] 
\in \mathbb{C}^{1 \times N_T}$, and $\mathbf{H}[m] \in \mathbb{C}^{N_I \times 
N_T }$ denote the channels between IRS-user $k$, BS-user $k$, and BS-IRS, 
respectively. Also, $\boldsymbol{\Phi}[n]=\text{diag}(\boldsymbol{\phi}[n]) \in 
\mathbb{C}^{N_I \times N_I }$ represents the phase shift matrix of the IRS with 
$N_I$ elements, and ${z}_{k}[m,n]\sim 
\mathcal{CN}(0,\sigma^{2})$\footnote{We consider 
that noise variances are the same, i.e., $\sigma_{j}^{2} = \sigma_{k}^{2} = 
\sigma^{2}$.} indicates the noise at receiver $k$.
 Substituting (\ref{txvector}) into (\ref{channeld}), we obtain 
 (\ref{urllckk}), see the top 
 of the next page, where
 $I_{\text{k,URLLC}}$ denotes interference at user $k$. In a similar fashion,
 the signal received at eavesdropper $j$ is given by
\begin{align}\label{evsdrp}
  \hspace{-3mm}y_{j}[m,n]=\bar{\mathbf{h}}^H_j[m]\mathbf{x}[m,n] + {z}_{j}[m,n],
 \end{align}
 where we have defined 
 $\bar{\mathbf{h}}^H_j[m]:=\mathbf{h}^{H}_{j}[m]\boldsymbol{\Phi}[n]
 \mathbf{H}[m]+\mathbf{g}^{H}_{j}[m]$. Upon substituting (\ref{txvector}) into 
 (\ref{evsdrp}), we obtain the received signal at 
 eavesdropper $j$ in (\ref{evsdrp_j}) on top of next page. The corresponding 
 channel vectors 
 are $\mathbf{h}^H_{j}[m] \in \mathbb{C}^{1 \times N_I}$ and 
 $\mathbf{g}^H_{j}[m] \in \mathbb{C}^{1 \times N_T}$ for the IRS-eavesdropper 
 $j$ and BS-eavesdropper $j$ and ${z}_{j}[m,n]\sim \mathcal{CN}(0,\sigma^{2})$ 
 is the noise at eavesdropper $j$. We use $\gamma_k[m,n]$ and 
 $\gamma_{jk}[m,n]$ to represent the SINR for user $k$ at the intended receiver 
 and eavesdropper $j$ respectively. 
 \begin{table*}
 	\begin{align}\label{urllckk}
 	y_{k}[m,n]&= \bar{\mathbf{h}}_{k}^H[m,n]\mathbf{w}_{k}[m,n]s_{k}[m,n] +  
 	\underbrace{\sum_{l\neq 
 	k}^{K}\bar{\mathbf{h}}_{k}^H[m,n]\mathbf{w}_{l}[m,n]s_{l}[m,n]}_{\substack 
 	{I_{\text{k,URLLC}}}} + \bar{\mathbf{h}}_{k}^H[m,n]\mathbf{v}[m,n] + 
 	{z}_{k}[m,n],
 	\end{align}
 \end{table*} 
 \begin{table*}
 	 \vspace{-8mm}
 \begin{align}\label{evsdrp_j}
 y_{j}[m,n]&= \bar{\mathbf{h}}_{j}^H[m,n]\mathbf{w}_{k}[m,n]s_{k}[m,n] +  
 \underbrace{\sum_{l\neq 
 k}^{K}\bar{\mathbf{h}}_{j}^H[m,n]\mathbf{w}_{l}[m,n]s_{l}[m,n]}_{\substack 
 {I_{\text{jk, URLLC}}}}  + \bar{\mathbf{h}}_{j}^H[m,n]\mathbf{v}[m,n] + 
 z_{j}[m,n],
 \end{align}
 \vspace{-9mm}
\end{table*}
      
\section{Problem Formulation}
By considering asymptotically long codewords, both error probability and 
information leakage can be made arbitrarily small as long as transmission rate 
is kept below the secrecy capacity \cite{Csiszar1}. Unfortunately, the long 
codeword assumption is not practical in URLLC applications. A closed-form 
achievable secrecy rate for short packet transmission (SPT) was derived for 
additive white Gaussian noise (AWGN) channel in \cite{wangsecrecy}, and later 
extended to the multi-carrier scenario \cite{ghanem2020resource}. 

If one desires a maximum packet error probability of $\epsilon_{k}$ at the 
intended user, and a maximum information leakage of $\delta_{j,k}$ from user 
$k$ to 
eavesdropper 
$j$, the total number of securely transmitted bits to user $k$ is 
derived by \cite{ghanem2020resourcesec}. Its multi-carrier extension is given by
\begin{eqnarray}
\bar{B}^{}_{k}  &=&   \bar{n} 
\sum_{m=1}^{M}\sum_{n=1}^{N} \log_{2}(1+\gamma_{k}[m,n])\label{norm2a_1}\\& 
&-  
aQ^{-1}(\epsilon_{k})
\left({\sum_{m=1}^{M}\sum_{n=1}^{N}\bar{n}\ 
	Z_{k}[m,n]}\right)^{\frac{1}{2}}\label{norm2a_2}\\ & &\hspace{-1cm}-\max_{j 
	\in 
	\{1,2,...,J\}} 
\Bigg(\bar{n}\sum_{m=1}^{M}\sum_{n=1}^{N} 
\log_{2}(1+\gamma_{j,k}[m,n])\label{norm2a_3}\\& &\hspace{0.75cm}+ 
aQ^{-1}(\delta_{j,k})
\left({\sum_{m=1}^{M}\sum_{n=1}^{N}\bar{n} Z_{j,k}[m,n]}\right)^{\frac{1}{2}} 
\Bigg),\nonumber
\end{eqnarray}
Channel dispersion is defined as  
$Z_{k}[m,n]=\big(1-{(1+\gamma_{k}[m,n])^{-2}}\big), \forall k$ and 
$Z_{j,k}[m,n]=\big(1-{(1+\gamma_{j,k}[m,n])^{-2}}\big), \forall j, k$ 
\cite{thesis}. In practice, we would like to guarantee a minimum QoS of 
$B_k^{\text{req}}$ to user $k$, which represents the minimum number of securely 
communicated bits that user $k$ demands. 

Let us define $\mathbf{w}:=\{\mathbf{w}_{k}[m,n], \forall k,m,n\}$, 
$\mathbf{V}:=\{\mathbf{V}[m,n], \forall m,n\}$, and 
$\mathbf{\Phi}:=\{\mathbf{\Phi}[n], \forall n\}$. Next, we formulate the 
resource 
allocation problem which aims to minimize the 
total transmit power at the BS while guaranteeing a minimum quality of service 
for each URLLC user. To this end, the main optimization problem is given by  
\begin{IEEEeqnarray}{lll}\label{P}
	&& \underset{\mathbf{\mathbf{w}, \mathbf{V}, \mathbf{\Phi}}}{ \text{min}}
	\,\,\
	\sum_{n=1}^{N}\sum_{m=1}^{M}\left(\sum_{k=1}^{K}\|\mathbf{w}_k[m,n]\|^2 + 
	\Tr\left(\mathbf{V}[m,n]\right)\right)\nonumber\\ &&\hspace{0.5cm}
	\mbox{s.t.}~~
	\bar{B}^{}_{k}\Big(\mathbf{w},\mathbf{V},{\mathbf{\Phi}}\Big) \geq 
	B_k^{\text{req}}, \quad \forall  k\nonumber\\ 
	&&\hspace{1.2cm}\mathbf{w}_k[m,n] 
	= 
	0, \quad \forall n > 
	D_k,\quad \forall k,\nonumber\\
	&& \hspace{1.2cm}\mathbf{V}[m,n] \succcurlyeq 0, \forall m,n,\:\:\:\:\: 
	|\mathbf{\Phi}_{i,i}(n)| = 1, \forall i, n, 
\end{IEEEeqnarray}
where $\mathbf{\Phi}_{i,i}[n]$ is the $i$-'th diagonal element of matrix 
$\mathbf{\Phi}[n]$ for $i =\{1,2,...,N_I\}$, and $D_k$ represents the maximum 
tolerable delay for user $k$. 

Optimization problem formulated in (\ref{P}) is non-convex with coupling 
between optimization variables $\mathbf{w}$,$\mathbf{V}$, and $\mathbf{\Phi}$ 
through the QoS constraint. To tackle these issues, we apply block coordinate 
descent (BCD) and utilize successive convex 
approximation (SCA) to iteratively solve each non-convex 
sub-problem. \vspace{-0.4cm}

\begin{table*}
	\begin{align}\label{sinrk}
	\gamma_{k}[m,n]=\frac{\Tr\Big(\tilde{\mathbf{\Phi}}[n] \mathbf{G}_{k}[m] 
		\mathbf{W}_k[m,n] \mathbf{G}_{k}^H[m]\Big)}{\sum_{{l}\neq k}^{K} 
		\Tr\Big(\tilde{\mathbf{\Phi}}[n] \mathbf{G}_{k}[m] \mathbf{W}_l[m,n] 
		\mathbf{G}_{k}^H[m]\Big)+\Tr\Big(\tilde{\mathbf{\Phi}}[n] 
		\mathbf{G}_{k}[m] \mathbf{V}[m,n] \mathbf{G}_{k}^H[m]\Big)+\sigma^2},
	\end{align}
\end{table*}
\begin{table*}
	\vspace{-5mm}
	\begin{align}\label{evsdr_sinr}
	\gamma_{jk}[m,n]=\frac{\Tr\Big(\tilde{\mathbf{\Phi}}[n] \mathbf{G}_{j}[m] 
		\mathbf{W}_k[m,n] \mathbf{G}_{j}^H[m]\Big)}{\sum_{{l}\neq k}^{K} 
		\Tr\Big(\tilde{\mathbf{\Phi}}[n] 
		\mathbf{G}_{j}[m] \mathbf{W}_l[m,n] 
		\mathbf{G}_{j}^H[m]\Big)+\Tr\Big(\tilde{\mathbf{\Phi}}[n] 
		\mathbf{G}_{j}[m] \mathbf{V}[m,n] \mathbf{G}_{j}^H[m]\Big)+\sigma^{2}
},
	\end{align}
\end{table*}

\section{Our Proposed BCD Approach}
To facilitate solving \eqref{P} via semi-definite program (SDP), we define 
positive semi-definite matrices $\mathbf{W}_k[m,n] := 
\mathbf{w}_k[m,n] \mathbf{w}_k[m,n]^H$, and we introduce 
$\mathbf{W}_k$ as collections of these $\mathbf{W}_k[m,n],~\forall n,m$. 
Furthermore, we define 
$\tilde{\mathbf{\Phi}}[n] := \tilde{\boldsymbol{\phi}}[n] \tilde{\boldsymbol 
{\phi}}^H[n],~\forall n$ where $\tilde{\boldsymbol{\phi}}[n] := 
[\boldsymbol{\phi}^H[n] , 1]^H $. The SINR definitions can be compactly 
written as traces. For instance, the 
numerator of $\gamma_k[m,n]$  can be written as
\begin{IEEEeqnarray}{lll}\label{channel}
	\Big|\mathbf{h}^{H}_{k}[m]\boldsymbol{\Phi}[n]\mathbf{H}[m] 
	\mathbf{w}_k[m,n] +\mathbf{g}^{H}_{k}[m] \mathbf{w}_k[m,n]\Big|^2 
	=\nonumber\\ \hspace{1.5cm}\Tr\left(\tilde{\mathbf{\Phi}}[n] 
	\mathbf{G}_{k}[m] 
	\mathbf{W}_k[m,n] \mathbf{G}_{k}^H[m]\right) , 
\end{IEEEeqnarray}
where $\mathbf{G}_{k}[m] = 
\Big[\Big(\text{diag}(\mathbf{h}^{H}_{k}[m])\mathbf{H}[m]\Big)^T 
\mathbf{g}^{*}_{k}[m]\Big]^T$. Subsequently, SINRs at intended users $k$ and 
eavesdropper $j$ are given by \eqref{sinrk} and \eqref{evsdr_sinr} respectively.
Next, we reformulate the QoS expression as
\begin{eqnarray*} 
	&&\bar{B}^{}_{k} = 
	R_k(\boldsymbol{\gamma}_k)- 
	C_k(\boldsymbol{\gamma}_k) -   \max_{j \in 
		\{1,2,...,J\} } C_{j,{k}}(\boldsymbol{\gamma}_{j,k}),
\end{eqnarray*}
where $R_k$, $C_k$, and $C_{j,k}$ are given by \eqref{norm2a_1}, 
\eqref{norm2a_2}, and \eqref{norm2a_3}, respectively. By defining slack 
variables $\tau_k:=\max_{j \in 
	\{1,2,...,J\} } C_{j,{k}}$ and auxiliary variable $\alpha_k[m,n]$, and  
$\zeta_{j,k}[m,n]$ to decouple the constraints, an equivalent optimization 
problem to \eqref{P} is 
formulated as
\begin{IEEEeqnarray}{lll}\label{P_bar}
&&  \hspace{-0.6cm}\underset{\mathbf{\mathbf{W}, \mathbf{V}, 
\tilde{\mathbf{\Phi}}, \boldsymbol{\tau}, \boldsymbol{\alpha}, 
\boldsymbol{\zeta}  }}{ \text{min}}
\,\,\,  \sum_{n=1}^{N}\sum_{m=1}^{M}\Big(\sum_{k=1}^{K}\Tr(\mathbf{W}_k[m,n]) + 
\Tr\big(\mathbf{V}[m,n]\big)\Big) \nonumber\\ ~\text{s.t.}~~
&& \mathrm{C1a} \: : R_k(\boldsymbol{\alpha}_k) -  C_k(\boldsymbol{\alpha}_k) - 
\tau_k  \geq B_k^{\text{req}}, \forall k,\nonumber\\
&& \mathrm{C1b} \: : \tau_k  \geq C_{j,{k}}(\boldsymbol{\zeta}_{j,k}) 
\:\:,\forall j,k,  \nonumber \\
&& \hspace{2mm}\mathrm{C}2: \Tr(\mathbf{W}_k[m,n]) = 0, \forall n > D_k, 
\forall k,\nonumber\\
&& \hspace{2mm}\mathrm{C}3: \mathbf{V}[m,n] \succcurlyeq 0, \forall 
m,n,~\mathrm{C}4: \mathbf{W}_k[m,n] \succcurlyeq 0, 
\forall 
k,m,n,\nonumber\\
&& \hspace{2mm}\mathrm{C}5: \Rank(\mathbf{W}_k[m,n]) \leq 1 , \forall k,m,n,\nonumber\\
&& \hspace{2mm}\mathrm{C}6: \:   \text{diag}(\tilde{\mathbf{\Phi}}[n]) = 
I_{N_{I}+1}, \forall n,\quad \mathrm{C}7: \:   
\tilde{\mathbf{\Phi}}[n] \succcurlyeq 0, \forall n,\nonumber\\
&& \hspace{2mm}\mathrm{C}8: \:   \Rank(\tilde{\mathbf{\Phi}}[n]) =1, \forall n,\nonumber\\
&& \hspace{2mm}\mathrm{C}9: \: {\alpha}_k[m,n] \leq \gamma_k[m,n] , \forall k,m,n,\nonumber\\
&& \hspace{2mm}\mathrm{C}10:\: \zeta_{j,k}[m,n] \geq \gamma_{j,k}[m,n] , \forall j,k,m,n,
\end{IEEEeqnarray}
where, $\boldsymbol{\tau}$, $\boldsymbol{\alpha}$, and $\boldsymbol{\zeta}$ are 
the collection of optimization variables $ {\tau_k} \forall k$, 
$\boldsymbol{\alpha_k}\forall k$, and $\boldsymbol{\zeta_{j,k}}\forall j,k$, 
respectively. For a single carrier system, i.e. $M=1$, 
\cite{ghanem2020resourcesec} has proven that the constraints $\mathrm{C}$9 and 
$\mathrm{C}$10 hold with 
equality at the optimum. For the multi-carrier setup, our numerical results 
indicate that they are tight at the achieved sub-optimal solution as well. 

Finally, we apply BCD to problem \eqref{P_bar} and decompose it into two 
sub-problems $\tilde{\text{P}}$1 and $\tilde{\text{P}}$2. They are given by
\begin{IEEEeqnarray}{lll}\label{ptilde1} \tilde{\text{P}}1 :
	&&\underset{\mathbf{\mathbf{W}, \mathbf{V}, \boldsymbol{\tau}, 
	\boldsymbol{\alpha}, \boldsymbol{\zeta}  }}{ \text{min}}
	\,\,\,  
	\sum_{n=1}^{N}\sum_{m=1}^{M}\Big(\sum_{k=1}^{K}\Tr(\mathbf{W}_k[m,n]) + 
	\Tr\big(\mathbf{V}[m,n]\big)\Big) \nonumber\\
	\text{s.t.} &&  
	\mathrm{C}\mathrm{1a},~\mathrm{C}\mathrm{1b},~\mathrm{C}2,~\mathrm{C}3,
	\mathrm{C}4,~\mathrm{C}5,~\mathrm{C}9,~\mathrm{C}10.
\end{IEEEeqnarray}
\begin{IEEEeqnarray}{lll}\label{Ptilde2}
	\hspace{-0.75cm}\tilde{\text{P}}2 :
	&&~~\underset{ \check{\mathbf{\Phi}}, 
	p}{\text{min}}\label{Ptilde2}
	\,\,\, p \nonumber\\
	\hspace{-0.75cm}\text{s.t.} \:\:
	&&\hspace{-0mm}\tilde{\mathrm{C}}6:\: \text{diag}(\check{\mathbf{\Phi}}[n]) 
	=  p I_{N_{I}+1}, \forall n,~  
	\mathrm{C}7,~\mathrm{C}8,~\mathrm{C}9,~\mathrm{C}10.
\end{IEEEeqnarray}
We have defined $\check{\mathbf{\Phi}}[n]:=p\tilde{\mathbf{\Phi}}[n]$. Specific 
formulation of the second sub-problem is attributed to \cite{yu2020irs}, where 
it is revealed that solving $\tilde{\text{P}}1$ and $\tilde{\text{P}}2$ 
iteratively yields a sequence of decreasing objective values in \eqref{P_bar}. 
The constraints $\mathrm{C}1$a, $\mathrm{C}1$b, 
$\mathrm{C}5$, $\mathrm{C}9$, 
and $\mathrm{C}10$ are non-convex in $\tilde{\text{P}}1$, while constraint 
$\mathrm{C}8$ is non-convex in 
$\tilde{\text{P}}2$. Next, we tackle most of these non-convex constraints via 
SCA. 

\section{SCA for BCD Sub-Problems}
To facilitate the application of SCA to $\tilde{\text{P}}1$, we employ a first 
order Taylor series approximation for $\mathrm{C}$1a, and $\mathrm{C}$1b. 
This leads to the following convex constraints:
\begin{IEEEeqnarray}{lll}\hspace{-2cm}
\mathrm{\bar{C}1a}: R_k(\boldsymbol{\alpha}_k) -  
\tilde{C}_k(\boldsymbol{\alpha}_k) - \tau_k  \geq B_k^{\text{req}},~~\forall k, 
\\
\hspace{-2cm}\mathrm{\bar{C}1b}: \tau_k  \geq 
\tilde{C}_{j,{k}}(\boldsymbol{\zeta_{j,k}}),~~\forall j,k, 
\end{IEEEeqnarray}
where $\tilde{C}_k(\boldsymbol{\alpha}_k) =  
{C}_k(\boldsymbol{\alpha}_{k}^{(i)}) +  
\left(\nabla_{\boldsymbol{\alpha}_{k}^{}}{C}_k\right)^T(\boldsymbol{\alpha}_k - 
\boldsymbol{\alpha}_{k}^{(i)})$ and 
$\tilde{C}_{j,{k}}(\boldsymbol{\zeta}_{j,k}) = 
{C}_{j,{k}}(\boldsymbol{\zeta}_{j,k}^{(i)}) + 
\left(\nabla_{\boldsymbol{\zeta}_{j,k}^{}}{C}_{j,{k}}\right)^T
(\boldsymbol{\zeta}_{j,k}^{}
 - 
\boldsymbol{\zeta}_{j,k}^{(i)})$. Here, $\boldsymbol{\alpha}_{k}^{(i)}$ and 
$\boldsymbol{\zeta}_{j,k}^{(i)}$ denote feasible points which are set equal to 
the optimum values from the previous SCA iteration $i$. Afterwards, we deal 
with 
the non-convex constraints $\mathrm{C}$9 and $\mathrm{C}$10. First, several 
auxiliary variables are introduced. Secondly, Schur complement is utilized to 
convert the convex constraints into linear matrix inequalities (LMIs). Thirdly, 
Taylor series expansion is exploited to approximate non-convex terms with an 
affine surrogate as part of the SCA procedure. Details can be found for the 
single carrier setup in \cite{ghanem2020resourcesec} and extension to 
multi-carrier setup is straightforward. Finally, the rank constraint in 
$\mathrm{C}$5 is dropped relaxing the problem. Thus, $\tilde{\text{P}}1$ could 
be sub-optimally solved via a series of convex SDPs via CVX. SCA decreases the 
objective at every iteration and is 
guaranteed to converge. To 
summarize, the SCA-based reformulation of  
$\tilde{\text{P}}1$ is solved efficiently by a straightforward multi-carrier 
extension of \cite[Algorithm 
1]{ghanem2020resourcesec}. 

Next, we focus on $\tilde{\text{P}}2$ in \eqref{Ptilde2} and optimize the phase 
shift matrix 
${\tilde{\mathbf{\Phi}}}$, while fixing the 
variables $\hat{\mathbf{w}}_k$, 
$\hat{\mathbf{V}}$, $ \hat{\boldsymbol{\tau}}, \hat{\boldsymbol{ {\alpha}}},$ 
and $\hat{\boldsymbol{ {\zeta}}}$ to their 
optimum obtained from the 
previous BCD step of solving $\tilde{\text{P}}$1. It should be 
mentioned that $\mathrm{C}$9 and $\mathrm{C}$10 are non-convex with respect to 
optimization variables in $\tilde{\text{P}}1$, while they are convex with 
respect to optimization variables in $\tilde{\text{P}}2$. Utilizing 
\eqref{sinrk}, $\mathrm{C}$9 is reformulated into a linear 
inequality with respect to $\check{\mathbf{\Phi}}$ as follows
\begin{IEEEeqnarray}{lll}\label{P12}
\mathrm{C}9: &&~~\hat{\alpha}_k(m,n)  \Bigg(\sum_{{l}\neq k}^{K} 
\Tr\Big(\mathbf{G}_{k}[m] 
\check{\mathbf{W}}_l[m,n]\mathbf{G}_{k}^H[m]
{\check{\mathbf{\Phi}}}[n]\Big) \nonumber\\&&\hspace{0.75cm}+  
\Tr\Big(\mathbf{G}_{k}[m] 
\check{\mathbf{V}}[m,n]\mathbf{G}_{k}^H[m]{\check{\mathbf{\Phi}}}[n]\Big)
 +\sigma^2\Bigg)  \nonumber\\&& \hspace{-0.5cm}\leq \Tr\Big(\mathbf{G}_{k}[m] 
\check{\mathbf{W}}_k[m,n] 
\mathbf{G}_{k}^H[m]{\check{\mathbf{\Phi}}}[n] \Big),~~ \forall k,m,n
\end{IEEEeqnarray}
where $\check{\mathbf{W}}_k[m,n]:=\hat{\mathbf{W}}_k[m,n]/\hat{p}$, 
$\check{\mathbf{V}}[m,n]:=\hat{\mathbf{V}}[m,n]/\hat{p}$. Furthermore, we have 
$\hat{p}:=\sum_{n=1}^N\sum_{m=1}^M\Big(\sum_{k=1}^K\text{Tr}(\hat{\mathbf{W}}_k[m,n])+
\text{Tr}(\hat{\mathbf{V}}[m,n])\Big)$. Similarly, utilizing 
\eqref{evsdr_sinr}, $\mathrm{C}$10 can be reformulated as in \eqref{P12} with 
$\hat{\alpha}_k(m,n), \mathbf{G}_{k}[m]$ replaced by 
$\hat{\zeta}_{j,k}(m,n),\mathbf{G}_{j}[m]$ respectively and the inequality 
direction reversed. 
The next task is to address the non-convex rank one constraint $\mathrm{C}$8. A 
novel method to deal with these types of constraints is provided by 
\cite{dai19}. Their approach replaces the rank constraint with a semi-definite 
constraint 
\begin{equation}\label{rank_sca}
\mathrm{\tilde{C}}8:~~r_n \mathbf{I}_{N_I} - 
\check{\mathbf{U}}_{N_I}^{(i)}[n]^H \check{\mathbf{\Phi}}[n] 
\check{\mathbf{U}}_{N_I}^{(i)}[n] \succcurlyeq 0.
\end{equation}
Here, $\check{\mathbf{U}}_{N_I}[n]$ represents the $(N_I+1)\times N_I$ matrix 
whose columns are the smallest $N_I$ eigenvectors of 
$\check{\mathbf{\Phi}}[n]$. In order for $\check{\mathbf{\Phi}}[n]$ to be rank 
one, $\mathrm{\tilde{C}}$8 should hold with $r_n=0$. Since 
$\check{\mathbf{U}}_{N_I}[n]$ is not available, we use SCA and utilize the 
smallest $N_I$ eigenvectors of $\check{\mathbf{\Phi}}^{(i)}[n]$, which is the 
optimum solution of previous SCA iteration and denote them by 
$\check{\mathbf{U}}_{N_I}^{(i)}[n]$ in \eqref{rank_sca}. Furthermore, to 
ensure that ultimately $r_n=0$ while obtaining an initial feasible point 
easily, we penalize $r_n$ in the objective. At SCA iteration $i$, 
the following convex optimization problem is solved
\begin{IEEEeqnarray}{lll}\label{P2-1}
	\underset{ \check{\mathbf{\Phi}}, p,\mathbf{r} }{\text{min}}
	\,\,\, ~~~ \Upsilon:= p + \lambda^{(i)} \sum_{n = 1}^N r_n  \\
	\text{s.t}\:\:\:
	\tilde{\mathrm{C}}6,~\mathrm{{C}}7 ,~ \mathrm{\tilde{C}}8, 
	~\mathrm{C}{9},~\mathrm{{C}}{10}, \nonumber
\end{IEEEeqnarray} 
where $\lambda^{(i)}$ represents a sequence of increasing weights. The 
proposed algorithm for the phase shift 
optimization is summarized in Algorithm 1. 
\begin{algorithm}[t]
	\caption{SCA for $\tilde{\text{P}}$2 in \eqref{Ptilde2}}
		\begin{enumerate}
		\item \textbf{Initialize} 
		$\check{\mathbf{\Phi}}^{(1)}$,~$\lambda^{(1)}$,~
		$\lambda_{{\max}}\gg 1$,~$\eta > 1$, and $0 
		\leq Er_\text{SCA} \ll 1 $. 
		\item \textbf{Repeat} 
		\item Solve  (\ref{P2-1}) for given $\check{\mathbf{\Phi}}^{(i)}$ to 
		obtain $\check{\mathbf{\Phi}}^{(i+1)}$ 
		\item Set $i \longrightarrow i+1$, update 
		$\lambda^{(i+1)}  = \min(\eta \lambda^{(i)}, \lambda_{{\max}})$
		\item \textbf{Untill}  
		$\frac{\Big|\Upsilon(\check{\mathbf{\Phi}}^{(i+1)}) - 
		\Upsilon(\check{\mathbf{\Phi}}^{(i)})\Big|}
	{\Big|\Upsilon(\check{\mathbf{\Phi}}^{(i)})\Big|}$ $\leq$ 
		$Er_\text{SCA}$
		\item \textbf{Output} $\mathbf{\Phi}^* = 
		\frac{\check{\mathbf{\Phi}}^{(i)}}{p^{(i)}}$   
	\end{enumerate}
\end{algorithm}In the end, the overall BCD algorithm is summarized in Algorithm 
2.
\begin{algorithm}[t]
	\caption{BCD Algorithm for Solving \eqref{P_bar}}
	\begin{enumerate}
		\item \textbf{Initialize} $\big\{\mathbf{w}^{(1) 
		},\mathbf{V}^{(1)},{\mathbf{\Phi}}^{(1)}\big\}$, and $0 \leq 
		Er_\text{BCD} \ll 1 $. 
		\item \textbf{Repeat} 
		\item For given ${\mathbf{\Phi}} = {\mathbf{\Phi}}^{(\mu)}$, solve 
		\eqref{ptilde1} via the multi-carrier extension of  
		\cite[Algorithm 1]{ghanem2020resourcesec} to obtain 
		$\mathbf{w}^{(\mu+1)},\mathbf{V}^{(\mu+1)}$
		\item Given $\mathbf{w}^{(\mu+1)}, \mathbf{V}^{(\mu+1)}$, solve 
		(\ref{P2-1}) via 
		$\textbf{Algorithm 1}$ to obtain ${\mathbf{\Phi}}^{(\mu+1)}$. Set $\mu 
		\longrightarrow \mu +1$ 
		\item \textbf{Till} ratio of improvement in objective $\leq$ 
		$Er_\text{BCD}$
		\item \textbf{Return} ${\mathbf{w}}^* = \mathbf{w}^{(\mu)}, 
		\mathbf{V}^* = \mathbf{V}^{(\mu)},{\mathbf{\Phi}}^* = 
		{\mathbf{\Phi}}^{(\mu)}$
	\end{enumerate}
\end{algorithm}

\section{Numerical Results}
All channels, i.e., BS-IRS, IRS-user, BS-user, BS-eavesdropper, and 
IRS-eavesdropper which we denote by $x$ are modeled as $g_x 
\times PL_{d_x}$, 
where $g_x$ represents the small scale fading, and
${PL_{d_x}} =\left(\frac{c}{4\pi f_c d_{\rm ref}}\right) 
\left(\frac{d_x}{d_{\rm ref}}\right)^{-\Gamma_x} b_x$ describes the path-loss 
dependent 
large-scale fading. The first term in path-loss stands for the loss at a 
reference 
distance $d_{\rm ref}=1$ meter and carrier frequency of $f_c$. The second term
is distance-dependent path loss with exponent $\Gamma_x$ and third term $b_x$ is shadowing/blockage of direct channels.
Table I summarizes the selected parameters. Note that both path-loss exponent 
and Rician factor vary depending on the type of the link $x$. Besides being 
uniformly located in disks of different radius as in Table I, authentic URLLC 
users and eavesdroppers maintain the same channel Rician factor and path loss 
exponent.
\begin{table}[t]
	\centering
	\vspace{8mm}
	\caption{System parameters.} 
	\label{tab:table}
	\renewcommand{\arraystretch}{1.4}
	\scalebox{0.6}{%
		\begin{tabular}{|c||c|} 
			\hline
			Cell radius: Eavesdroppers, users & $r_{Ie} =50$ meters, $r_{Iu} =5$ meters \\ \hline
			Number and bandwidth of subcarriers, and time slots & $M = 32$, $\text{Bw} = 240$ kHz and $N = 4$ \\ \hline
			Carrier frequency and Noise power density  & $f_c = 6$ GHz and $ N_0$ = -174 dBm/Hz \\ \hline
			Number of bits per packet and system delay & $B_k^{\text{req}} =160$ bits and $T_f = 0.21667$ ms \\ \hline  
			Maximum base station transmit power $P_{\text{max}}$  &  $45$~dBm \\ \hline  
			Max error probability and information leakage  & $\epsilon_k = 10^{-6}$,  $\delta_{j,k} = 10^{-6}, \forall j,k$  \\ \hline 
			Racian factor  &  $K_\text{BI} = 10$, $K_\text{Bu} = 0$, and $K_\text{Iu} = 0$ \\ \hline
			Path loss exponent  &  $\Gamma_\text{BI} = 2.1$, $\Gamma_\text{Bu} = 3.5$, and $\Gamma_\text{Iu} = 2.1$ \\ \hline
	    	shadowing/blockage  &  $b_\text{Bu} = -10 $\:dB and $b_\text{Be} = -10 $\:dB, \\ \hline
	\end{tabular}}
	\vspace*{-5mm}
\end{table} 

The number of URLLC users and eavesdroppers are set to $K = 2$ and $J=2$  
respectively. Furthermore, $N_T$ = 2 and $N_I$ = 50. Our simulation geometry is 
according to Fig. \ref{model}, where users and eavesdropper are located in 
separate disks with radii specified in Table I. Furthermore, we consider the 
network center, BS position, and IRS location to be at $(0,0)$, $(0,-100)$, and 
$(50, 0)$, respectively. The distance between URLLC users/eavesdroppers disk 
centers from IRS are given by $d_\text{Iu}=4$ 
meters and $d_\text{Ie} = 200$ meters, respectively. The distance of users and 
eavesdroppers disk centers from BS are given by $d_\text{Bu} = 500$ meters and 
$d_\text{Be} = 505$ meters, respectively. In addition, we assume $D_1 = 2,$ 
and $D_k = 4, \forall k>1$ as delay requirement of users. The parameters of 
first BCD sub-problem, i.e.,
\cite[Algorithm 1]{ghanem2020resourcesec}, are set to  
$\{t=10, 
t_\text{max} = 10^6, \eta = 6, I_\text{max} = 16 \}$, while parameters of 
second BCD sub-problem, i.e., 
$\textbf{Algorithm 1}$ in this work, are set to $\{\lambda^{(1)}=0.1, 
\lambda_\text{max} = 10^5, \eta = 1.2, E_{r_\text{SCA}} = 10^{-5} \}$. We have 
defined $I_\text{max}$ as the maximum number of 
iterations that can be afforded.

A second, more practical scenario is also investigated where the channel of 
legitimate users and eavesdroppers are spatially correlated and eavesdroppers 
are closer to the BS with $d_\text{Be} = 250$. To model correlations, we 
exploit a spatial correlation matrix 
$\mathbf{R}$ to generate $\mathbf{g}_k ,\forall k$ and $\mathbf{g}_j ,\forall 
j$, while the other channels remain independent and unchanged. It is assumed 
that $[\mathbf{R}]_{i,j} = \rho^{|i-j|}$ with $\rho = 0.95$ 
\cite{cui2019secure}.  
\subsection{Benchmark Schemes}
	\begin{itemize}
	\setlength{\itemsep}{1pt} 
	     \item {\textbf{SC}}: Secrecy capacity for infinite 
	     block length where all channel dispersions 
	     are omitted from constraints $\mathrm{C}\mathrm{1a}$ and 
	     $\mathrm{C}\mathrm{1b}$ in (\ref{P_bar}). This amounts to letting  
	     $\bar{n}$ go to infinity. The same BCD Algorithm 2 is utilized to 
	     find a sub-optimal solution for this scheme. It provides a lower bound 
	     on the total transmit power at the BS for FBL
	     \cite{ghanem2020resourcesec}.
	     
     \item {\textbf{Baseline 1}}: We adopt random phase shifts at the 
     IRS. Given a random phase shift matrix, we jointly optimize the 
     beamformers and AN at the BS via \cite[Algorithm 
     1]{ghanem2020resourcesec}.     
     \item {\textbf{Baseline 2}}: We consider conventional secure-URLLC 
     with No IRS and optimize the beamforming vector and AN at the BS 
     \cite{ghanem2020joint}.  
\end{itemize}

\vspace*{-1mm}
  \subsection{Simulation Results}
\begin{figure}[t]
  	\centering
  	\includegraphics[width=0.45\textwidth]{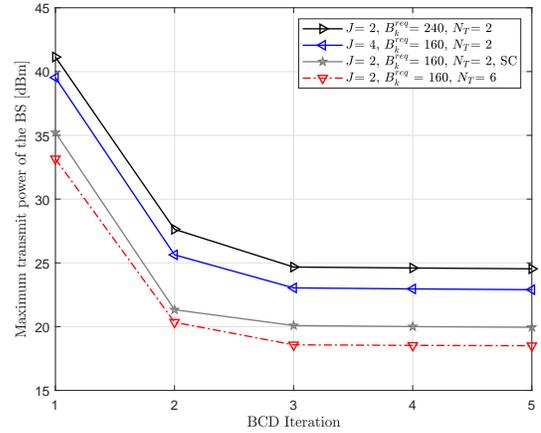}
  	\caption{Convergence speed of the proposed algorithm}
  	\label{fig20}
  	\vspace{-0.5cm}
  \end{figure}

  Fig. \ref{fig20} corroborates the fast convergence rate of the BCD 
  algorithm in approximately 5 iterations. This convergence occurs regardless 
  of the number of eavesdroppers, BS antennas, and required QoS, which is 
  suitable for URLLC use cases.
   \par
\begin{figure}[t]
   	\centering
   	   	\includegraphics[width=0.5\textwidth]{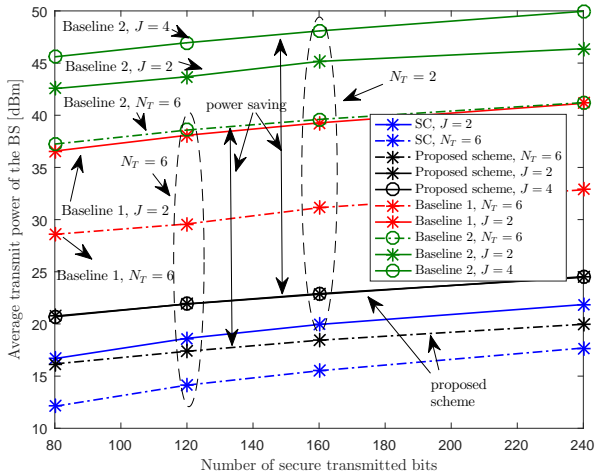}
   	\caption{Average transmit power versus number of secure bits per 
packet}
   	\label{fig2}
   	\vspace{-0.5cm}
   \end{figure}

 In Fig. \ref{fig2}, we study the impact of required number of secure 
   communication bits $B_k^{\text{req}}, \forall k$, on the average 
   transmit power at the BS for $N_T=2, 6$. It is evident that without IRS, BS 
   could only guarantee the required number of securely 
   transmitted bits at an exorbitant increase in its total 
   transmit power. Interestingly, 
   Baseline 1 outperforms Baseline 2 even though it exploits the IRS in a naive 
   way. Our 
   proposed 
   BCD enjoys substantial power savings versus both baselines. As 
   expected, SC lower 
   bound achieves the highest power saving. However, 
   SC is designed for infinite block length and not applicable to URLLC 
   scenarios. Finally, IRS ensures that even with small number 
   of active antennas, i.e., $N_T = 2$ or $N_T = 6$ one can still obtain 
   significant 
   power savings at the BS side.
   
   From another aspect, Fig. \ref{fig2} also illustrates the impact of number of
   eavesdroppers on  
   performance. When eavesdroppers outnumber the BS antennas, i.e., $N_T=2 < 
   J=4$, transmit beamforming at the BS would suffer from 
   insufficient spatial DoF for signal suppression in the direction of 
   eavesdroppers. This drawback is illustrated by Baseline 2 which yields an 
   excessive increase in BS transmit power. The presence of an IRS prevents 
   such a power 
   increase at the BS and enables the system to achieve the required secrecy 
   rate. Even when secrecy rate decreases by the presence of more 
   eavesdroppers, our proposed method can re-establish the needed QoS without 
   any noticeable increase in power, while this is not the case for no IRS. 
    
\begin{figure}[t!]
	\centering
	   		\includegraphics[width=0.5\textwidth]{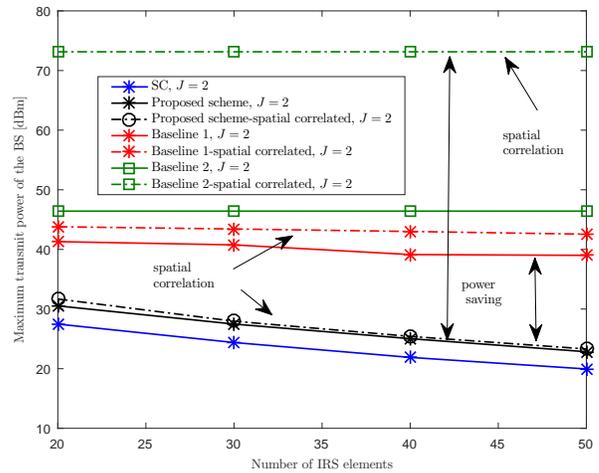}
	\caption{ Average transmit power versus number of IRS elements}
	\label{fig3}
	\vspace{-0.5cm}
\end{figure}

    In Fig. \ref{fig3}, we investigate the impact of spatially correlated 
    channels on average transmit power at the BS versus number of IRS elements. 
    One observes that Baseline 2 suffers a significant increase of BS 
    transmit power in spatially correlated channels in comparison to its 
    uncorrelated counterpart. This indicates that conventional techniques such 
    as BS beamforming and/or AN introduction at BS could not 
    achieve the required secrecy rate with a practically feasible BS power. In 
    contrast, the proposed scheme is robust to 
    spatially correlated BS-users/BS-eavesdroppers channels as well as stronger 
    BS-eavesdropper channel gains and the increase in BS transmit power is 
    hardly noticeable. This advantage comes from the extra DoFs appearing due 
    to IRS deployment which manages to realize constructive and 
    destructive combinations of the desired signal at legitimate users and 
    eavesdroppers, respectively. In addition, we observe that transmit power of 
    the proposed 
    scheme 
    decreases monotonically as the number of IRS elements increases even in 
    unfavorable channel conditions. While Baseline 1 
    avoids the significant power increase of Baseline 2, it still demands 
    significantly more power compared to the proposed scheme. Interestingly, 
    our proposed approach that considers finite block length limitations comes 
    surprisingly close to the unachievable SC benchmark.     

 \section{Conclusion}
Resource allocation for secure multiuser downlink IRS-enabled MISO-URLLC 
systems was investigated. To guarantee a given secrecy rate QoS in the finite 
block length regime, a non-convex optimization problem with the aim of 
minimizing the total BS transmit power was formulated. An efficient combination 
of BCD and SCA techniques were proposed to jointly design the BS beamformers 
and AN and IRS phase shifts. The proposed approach converges and can reach a 
sub-optimal solution of the main problem. Simulation results corroborated the 
improved performance achieved regardless of the channel conditions and 
increased 
robustness to number of eavesdroppers. 
\bibliography{ref}  
\bibliographystyle{IEEEtran}
\end{document}